# Multiple-partition cross-modulation programmable metasurface empowering wireless communications

Jun Wei Zhang[1,2,†], Zhen Jie Qi[1,2,†], Li Jie Wu[1,2], Wan Wan Cao[1,2], Xinxin Gao[3], Zhi Hui Fu[1,2], Jing Yu Chen[1,2], Jie Ming Lv[1,2], Zheng Xing Wang[3], Si Ran Wang[3], Jun Wei Wu[1,2], Zhen Zhang[4,*], Jia Nan Zhang[1,2,*], Hui Dong Li [1,2], Jun Yan Dai[1,2], Qiang Cheng[1,2,*], and Tie Jun Cui[1,2,*]

1 Institute of Electromagnetic Space, Southeast University, Nanjing 210096, China;
2 State Key Laboratory of Millimeter Waves, Southeast University, Nanjing 210096, China
3 State Key Laboratory of Terahertz and Millimeter Waves, City University of Hong Kong, Hong Kong, 999077, China
4 School of Electronics and Communication Engineering, Guangzhou University, Guangzhou 518055, China
†Equally contributed to this work.
*Email: zhangzhen@gzhu.edu.cn, jiananzhang@seu.edu.cn, qiangcheng@seu.edu.cn, and tjcui@seu.edu.cn

**ABSTRACT**

With the versatile manipulation capability, programmable metasurfaces are rapidly advancing in their intelligence, integration, and commercialization levels. However, as the programmable metasurfaces scale up, their control configuration becomes increasingly complicated, posing significant challenges and limitations. Here, we propose a multiple-partition cross-modulation (MPCM) programmable metasurface to enhance the wireless communication coverage with low hardware complexity. We firstly propose an innovative encoding scheme to multiply the control voltage vectors of row-column crossing, achieving high beamforming precision in free space while maintaining low control hardware complexity and reducing memory requirements for coding sequences. We then design and fabricate an MPCM programmable metasurface to confirm the effectiveness of the proposed encoding scheme. The simulated and experimental results show good agreements with the theoretically calculated outcomes in beam scanning across the E and H planes and in free-space beam pointing. The MPCM programmable metasurface offers strong flexibility and low complexity by allowing various numbers and combinations of partition items in modulation methods, catering to diverse precision demands in various scenarios. We demonstrate the performance of MPCM programmable metasurface in a realistic indoor setting, where the transmissions of videos to specific receiver positions are successfully achieved, surpassing the capabilities of traditional programmable metasurfaces. We believe that the proposed programmable metasurface has great potentials in significantly empowering the wireless communications while addressing the challenges associated with the programmable metasurface's design and implementation.

**Keywords:** Programmable metasurfaces, encoding scheme, cross modulation, multiple partition, beamforming, wireless communication system.

# INTRODUCTION

Metamaterial is composed of artificial subwavelength structures, which has been the subject of research for nearly two decades[1]. Numerous exciting devices and phenomena have emerged, enabling the flexible manipulation of electromagnetic (EM) waves and complementing or replacing conventional EM devices [2,3]. For greater convenience and ease of use, metasurfaces are garnering attention from researchers in both physics and information communities owing to their unique features of ultra-thin thickness, low loss, and ease of fabrication. They can be used to manipulate the incident EM waves with amplitude, phase, polarization, and frequency modulations using various artificial structures[4-12]. Based on these properties, many significant devices and interesting phenomena have emerged in various fields, including frequency-selective surfaces, filters, absorbers, EM cloaks, holograms, EM computing, EM information theory, and wireless communications[2,13-18].

Metasurfaces have evolved through passive metasurfaces to programmable metasurfaces. Passive metasurfaces are designed with specific metal patterns and materials, whose functions are fixed upon the completion of designs[9,19-24]. Therefore, they suffer from the shortcomings of inflexibility and uncontrollability. Programmable metasurfaces, developed from digital coding metasurfaces, can bridge the gap between the physics and information worlds[6,7,10,25-31]. These metasurfaces are integrated with tunable devices such as positive-intrinsic-negative (PIN) diodes, varactors, liquid crystal, and micro-electro-mechanical systems (MEMS). Tuning the operational states of these devices will alter the EM responses of metasurfaces, enabling the design of multifunctional metasurfaces. In a 1-bit phase-modulation metasurface integrated with PIN diodes, the phase response of a meta-element can be controlled by switching the PIN diode on and off. When the phases have opposite responses, the meta-element will be encoded as '0' and '1' states. Arranging the coding sequences or patterns of the '0' and '1' states on the meta-array enables the realization of various functionalities such as beam scanning, beam splitting, multi-beam, and radar cross section (RCS) reduction, thereby reconfiguring the EM environment[12,14,16].

Despite their powerful ability to manipulate EM waves, the programmable metasurfaces need additional modulation structures for controlling the states of tuning devices compared to the passive metasurfaces, resulting in an increased hardware complexity. Especially, as the scale of metasurface and number of tunable devices increase, deploying the control structures for these devices becomes increasingly challenging. The conventional modulation structures for programmable metasurfaces primarily consist of two main types. The first type involves meta-elements located in the same row or column sharing a common control signal[25,31-33]. This deployment method greatly simplifies the modulation structures and reduces the number of control ports. However, it also significantly reduces the capability to manipulate EM waves. For example, it is limited to scan the beam on one side and fails to achieve the beamforming in the whole space. Another modulation structure, an independent control type, provides control signals to the meta-elements individually[27,28,30,34,35]. This structure can competently achieve beamforming in free space with high precision. However, the number of control ports equals the number of tuning devices, exacerbating the pressure on the control terminal and increasing the burdens of modulation structures placement and simulation verification.

In many scenarios, hindered by the high complexity and cost of hardware structures, it is difficult for metasurfaces to perform high-precision beamforming in free space. Therefore, it is necessary to develop novel modulation schemes that can achieve the best trade-off between hardware cost and beamforming accuracy. In the past a few years, several research activities have been carried out to address the above issues. For example, the perimeter-controlled approach is proposed to simplify the biasing architecture by enabling column-row addressing of elements, thus realizing the beam scanning in free space[29]. However, this approach suffers from reduced directivity and deviations from the desired steering angle. A modulo-addition operation is proposed to enable terahertz programmable metasurfaces for beam steering in free space[26]. The cross-modulation structures are employed to control the polarization of liquid crystal, thereby tuning the EM responses with fewer control ports and simpler modulation structures. However, the persistent presence of multi-beams and limited control rate constrain the range of applications. A voltage control component is designed to modulate the devices in turn, which could control meta-elements independently with simplified structures[28]. However, the additional control devices on each meta-element increases the complexity of metasurface designs. In summary, the methods described above simplify the architecture of the modulation and enhance programmable metasurfaces' ability of beamforming in different ways, but the existing research literature lacks novel modulation schemes that take into account modulation structural complexity, manipulation rate, and pointing accuracy, particularly for diode-based programmable metasurfaces.

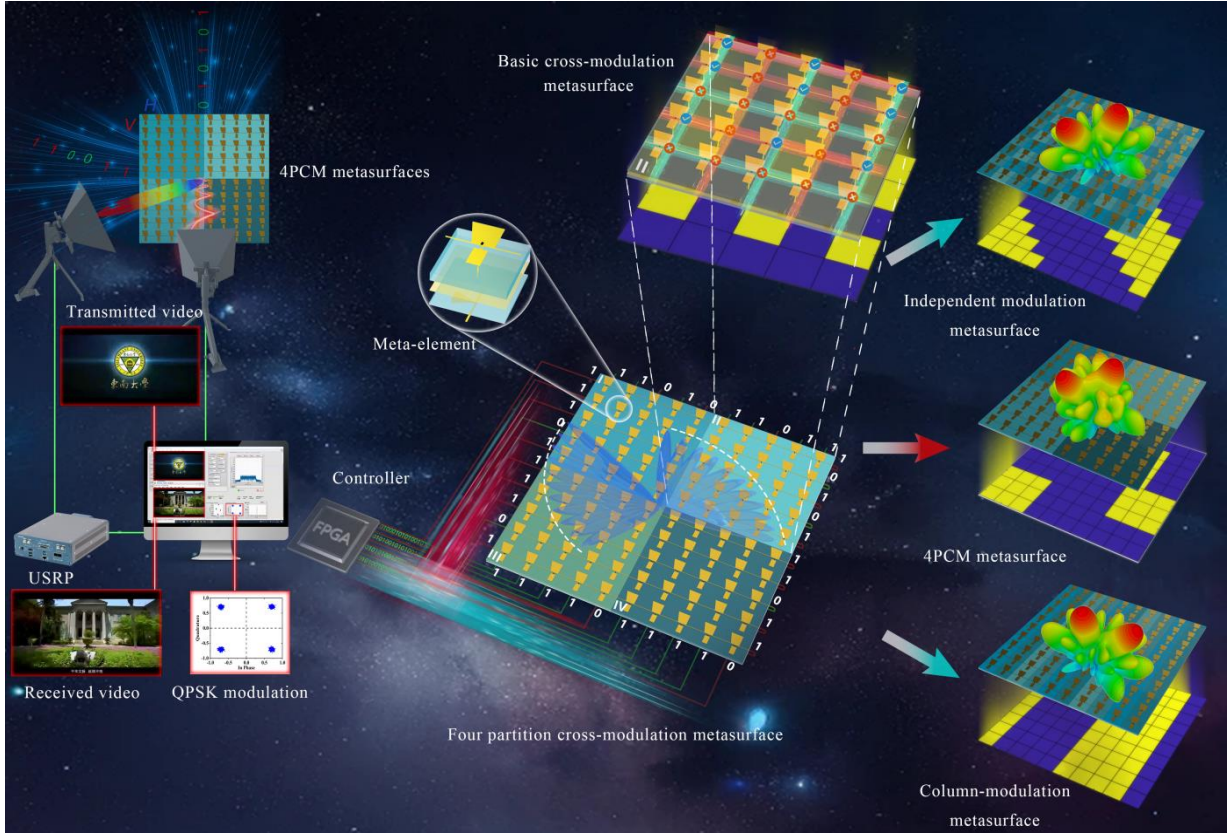

**Fig. 1 Illustration of the structure and working mechanism of four partition cross-modulation (4PCM) metasurfaces.** The new type of wireless communication system with the assistance of 4PCM metasurface (on the left side). The comparison of the three module methods in the complexity and the pointing accuracy of programmable metasurfaces (on the right side).

In this work, we propose a novel modulation structure for programmable metasurfaces with the multiple-partition cross-modulation (MPCM) method, as shown in Fig. 1. The precision and coverage of the beam pointing angles can be adjusted according to the partition items defined in the proposed modulation method. This proposed method aims to achieve a better trade-off between the beamforming accuracy and hardware complexity, thus promoting the implementation of programmable metasurfaces in the future wireless communications. We term the programmable metasurface driven by this proposed modulation method as MPCM programmable metasurface, which features much simpler control structures than the traditional counterpart. Subsequently, we develop an adaptive optimization algorithm to obtain the coding sequences that point at the desired directions in free space for the MPCM metasurfaces.

Following that, we design a programmable metasurface with four-partition cross-modulation (4PCM) structures and validate its capability of beam scanning in the H and E planes by theoretical calculations, full-wave simulations, and experiments. Furthermore, we investigate the relationship between the partition items of cross-modulation structures and the precision of beam pointing, as well as the level of side lobes. The theoretically calculations, full-wave simulations, and experimental results exhibit good consistency, verifying the effectiveness of the proposed method. Additionally, we build a realistic in-door scenario to demonstrate the performance of the fabricated 4PCM metasurface in the wireless communication assistance compared to the traditional programmable metasurfaces. The results from the demonstration illustrate the advantages of the MPCM metasurfaces in empowering wireless communication coverage. The proposed MPCM metasurfaces have great potential to alleviate the hardware complexity while maintaining a high beamforming precision in free space, particularly for the large-scale diode-based programmable metasurfaces. Due to such marvelous features, we envision this new type of programmable metasurfaces as a promising vehicle to promote the development and implementation of future wireless communication and radar detection systems.

## RESULTS

### Mechanism of the MPCM metasurfaces

The proposed MPCM programmable metasurfaces allow various numbers and combinations of partition items to offer strong flexibility and low hardware complexity, catering to diverse precision demands across various scenarios. The smallest number of partition items of the proposed MPCM programmable metasurfaces is one, named basic cross-modulation (BCM) metasurfaces or one partition cross-modulation (1PCM) metasurfaces, whose basic structure is shown in Fig. 2(a). The upper part is individually connected to the positive terminals of the PIN diodes in the same rows, while the lower part is connected to the negative terminals of diodes through the vias in the same columns. The high or low voltages are provided to the two parts to control the working states of the PIN diodes. There are four voltage combinations for tuning the diodes. The diode is at the on state when the row is high voltage and the column is low voltage. In the other three voltage combinations, the diode is in the off state, as shown in Fig. 2(b). Herein, after determining the voltages' states of all rows and columns, the diodes' working states in the metasurface array are determined.

For a BCM metasurface with the scale of $M \times N$, let the vectors $V$ with $M$ elements and $H$ with $N$ elements represent the feeding voltages of all rows and columns, respectively. Further let $V^i = 1$ and $V^i = 0$ ($i = 0, 1, \ldots, M$) represent high and low voltages at the $i$ row, respectively. Similarly, let $H^j = 0$ and $H^j = 1$ ($j = 0, 1, \ldots, N$) represent high and low voltage at the $j$ column, respectively. According to the two vectors defined above, we can obtain matrix $C$

$$C = V^T \cdot H, \qquad (1)$$

where $C$ represents the working states of the diodes in the metasurface array. The element $C^{(m,n)}$ ($m = 1,2,\ldots, M$ and $n = 1,2,\ldots, N$) represents the working state of the PIN diode located at the $m$-th row and $n$-th column, where $C^{(m,n)} = 0$ and $C^{(m,n)} = 1$ represent the off and on states of the $(m, n)$-th diode, respectively (refer to Supplementary Note 1 for more details). We illustrate the process of computing the coding sequence with vectors $V$ and $H$ in Fig. 2(c). Generally, the digital coding metasurfaces employ the codes "0" and "1" to represent the opposite EM phase responses (0 and $\pi$). Here, the codes "0" and "1" correspond to the off and on states of diodes, respectively. Therefore, the phase profile of BCM metasurfaces can be represented by the matrix $\pi C$.

For an MPCM metasurface composed of $K$-independent sub-BCM structures, we assume that the number of partition items is $K$, i.e., $K$ partition cross-modulation (KPCM) metasurfaces.

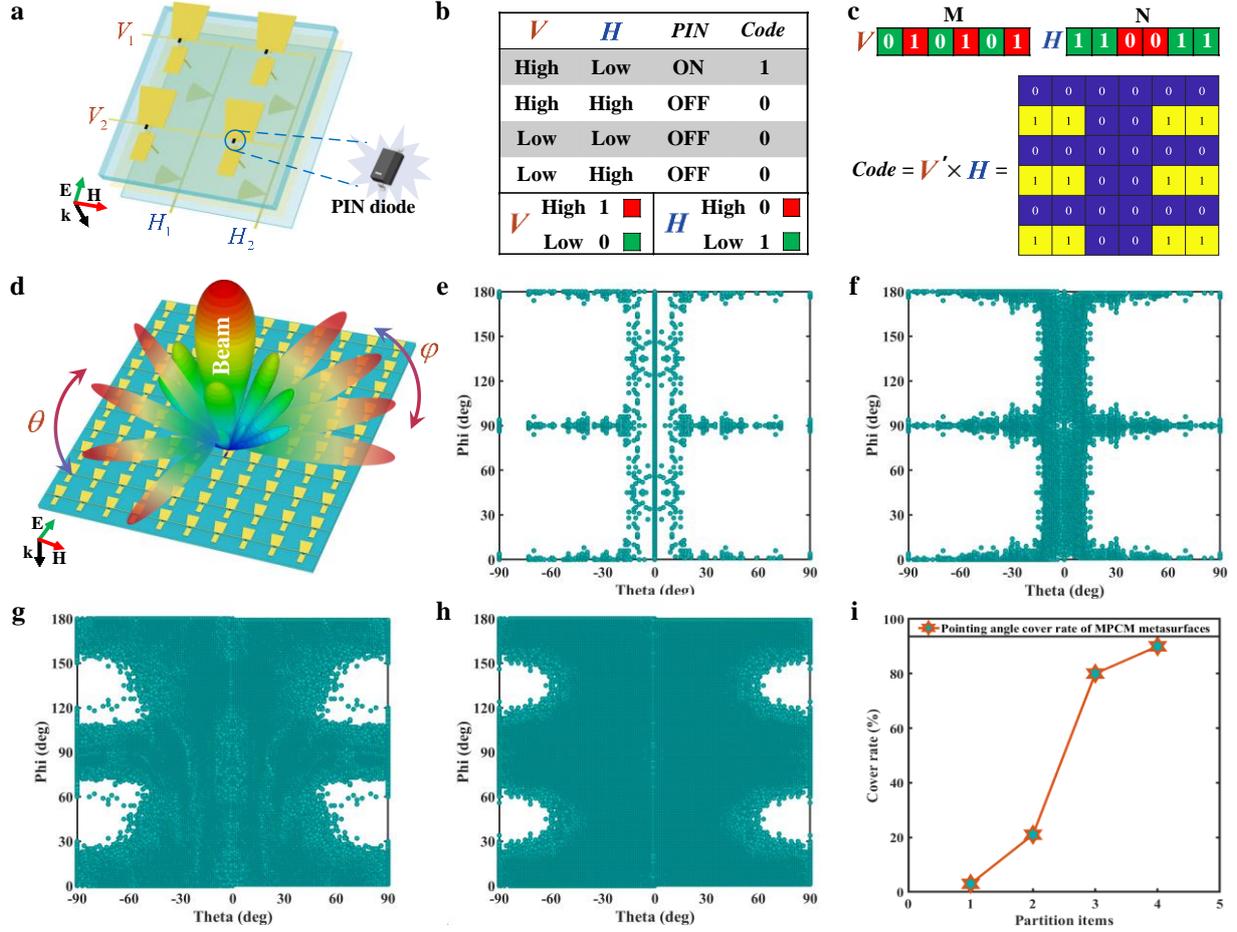

**Fig. 2 The encoding rules and beam pointing coverage performance of the proposed MPCM metasurfaces.** (a) The view of basic cross-modulation (BCM) structure. (b) The PIN diodes' working states of BCM structure and the control signal codes of vertical and horizontal feed lines. (c) The coding sequences representation rules of BCM metasurfaces. (d) Demonstration to the beam scanning with a pointing angle ($\theta$, $\varphi$) in the upper-half space by calculating the coding sequences of MPCM metasurfaces. (e) The far-field scattering beam pointing coverage of one partition cross-modulation (1PCM) with the scale of 6×6. (f) 2PCM. (g) 3PCM. (h) 4PCM. (i) Pointing angle coverage rate of MPCM metasurfaces with different partition items.

The guidance of sub-BCM structures partitioning strategies is described in Supplementary Note 2. Accordingly, $V_1, V_2, \ldots, V_K$ and $H_1, H_2, \ldots, H_K$ represent the vertical and horizontal control voltages' codes of the $K$ parts, respectively. Through the two group vectors and matrix operations, the coding sequences $C$ of KPCM metasurfaces (denoted by $C_{KPCM}$) can be represented by

$$C_{KPCM} = \begin{bmatrix} V_1^T \cdot H_1 & V_2^T \cdot H_2 & \cdots \\ \vdots & \vdots & \vdots \\ \cdots & V_{K-1}^T \cdot H_{K-1} & V_K^T \cdot H_K \end{bmatrix}. \qquad (2)$$

## Beam coverage ability analysis of the MPCM metasurfaces

Based on the above encoding rules, one can observe that the number of coding sequences increases as the number of partition items increases, resulting in more various far-field scattering patterns. For programmable metasurfaces comprising $M \times N$ periodically arranged meta-elements, their phase profiles determine the beam deflection angles, as shown in Fig. 2(d). Assuming that the reflection phase of the ($m, n$)-th element in the meta-array is $\phi(m, n)$. When plane waves normally illuminate the metasurface, the far-field scattering pattern in the Cartesian coordinate

can be described by [36]

$$F(\theta,\varphi) = \sum_{m=1}^{M}\sum_{n=1}^{N} E(\theta,\varphi) \cdot A(m,n) \cdot \exp\left(-i\begin{pmatrix}\phi(m,n)\\+(n-1)\cdot k\cdot dx\cdot \sin(\theta)\cdot \cos(\varphi)\\+(m-1)\cdot k\cdot dy\cdot \sin(\theta)\cdot \sin(\varphi)\end{pmatrix}\right), \qquad (3)$$

where $\theta$ and $\varphi$ represent the angles of elevation and azimuth, respectively. $E(\theta,\varphi)$ is the far-field pattern of the meta-elements. $k$ is the wave number in free space, and $dx$ and $dy$ are the side lengths of the meta-elements along the $x$- and $y$-axes, respectively. Based on equation (3), the scattering patterns and main-lobe direction of metasurfaces can be calculated once the coding sequence is fixed. According to the convolution operations of coding metasurfaces, the relation between the scattering patterns and phase distribution ($\pi C$) of metasurface can be expressed by

$$F(\theta,\varphi) = G(u,v) = \mathcal{F}\{Ae^{j\pi C}\}, \qquad (4)$$

where $u = \sin\theta\cos\phi$, $v = \sin\theta\sin\phi$. The $u$ and $v$ coordinates can express the scattering patterns of metasurfaces clearly[19].

To investigate the coverage ability of the MPCM metasurfaces, we adopt a 6×6 periodic meta-array to traverse all possible coding sequences with different numbers of partition items. In specific, for the BCM metasurface, the number of coding sequences is $2^{12}$. The 2PCM, 3PCM, and 4PCM metasurfaces have $2^{18}$, $2^{21}$, and $2^{24}$ coding sequences, respectively. The strategies of partitioning and the size of sub-BCM determination are shown in Supplementary Note 2. The main-lobe direction coverage performances of MPCM metasurfaces are shown in Figs. 2(e)-(h) corresponding to 1PCM, 2PCM, 3PCM, and 4PCM metasurfaces, respectively. The specific coverage rate is depicted in Fig. 2(i). It can be easily observed that as the number of partition items increases, the coverage range of the main lobe beam in the upper half-space expands.

We then develop a coding sequences acquisition algorithm according to classic array synthesis theories and new regime encoding rules to achieve the target beam deflection of MPCM metasurfaces. Assuming that ($\varphi_0$, $\theta_0$) is the main-lobe deflection angle, the phase distribution of the meta-element in the array corresponding to the main-lobe direction angle can be expressed as

$$\Phi(m,n) = -k\left(x_i \sin\theta_0 \cos\varphi_0 + y_i \sin\theta_0 \sin\varphi_0\right), \qquad (5)$$

where $\Phi(m, n)$ is the phase of the $(m, n)$-th meta-element and $(x_i, y_i)$ represents its position in the Cartesian coordinate system.

We define $X = [V_1, V_2, \ldots, V_K, H_1, H_2, \ldots, H_K]$, which includes all control voltages' codes of row-column crossings. The distribution reflection phases of MPCM metasurfaces can be described as $\pi C_{KPCM}$, calculated by equation (2), which is a function of $X$. Assuming that the beam deflection angle is ($\varphi_0$, $\theta_0$). We define the corresponding phase distribution of the programmable metasurfaces as $\Phi$, calculated by equation (5). Then we can define the objective function for the coding sequences optimization as follows

$$U_1(X) = \frac{\|\pi C_{KPCM}(X) - \Phi\|_F}{M \cdot N}, \qquad (6)$$

where $U_1$ represents the objective function evaluated at given values of $X$, $\|\cdot\|_F$ represents the Frobenius norm of a given matrix. The symbol $\Phi$ is used here to assist the metasurface in achieving the desired main-lobe direction angles. Based on equation (6), the optimization problem can be formulated by

$$X^* = \arg\min_{X} U_1(X), \qquad (7)$$

where $X^*$ represents the optimized solution of $X$. In this optimization process, we adopt the standard genetic algorithm to optimize the control voltages of the rows and columns and further

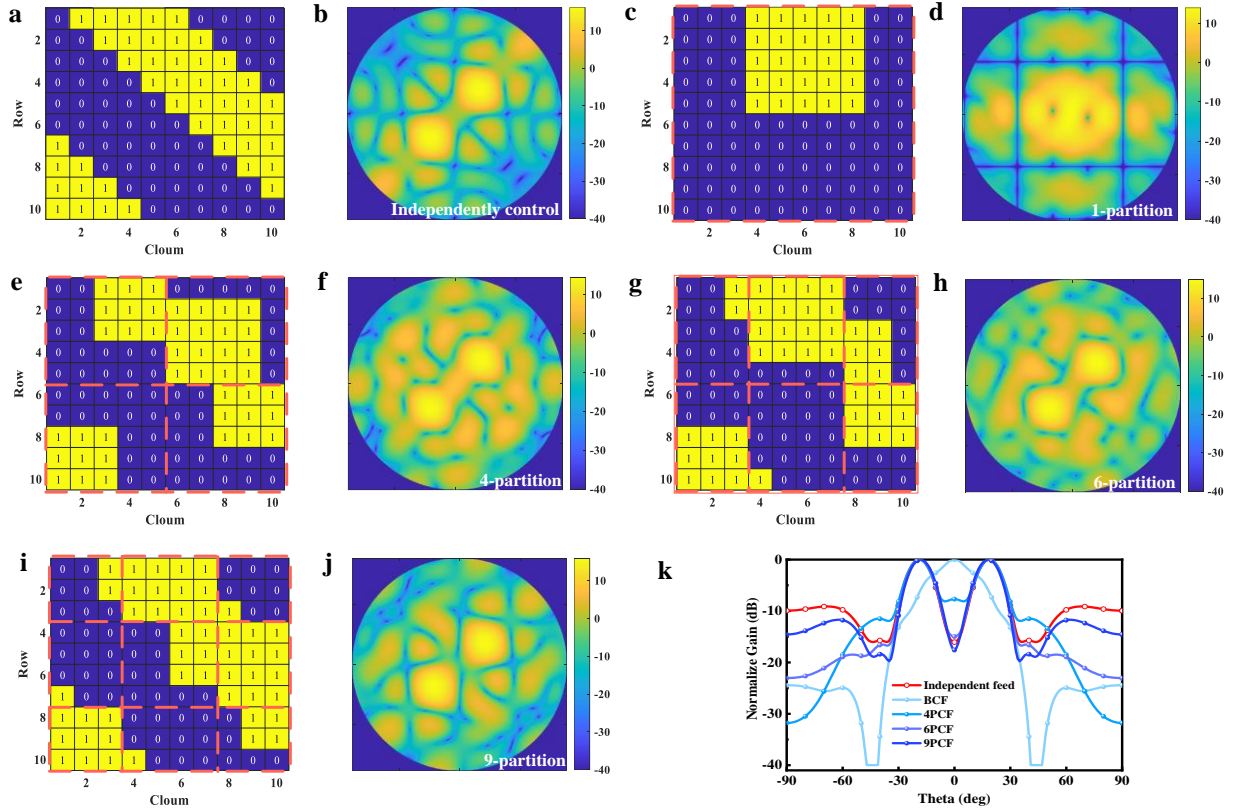

**Fig. 3 The coding sequences and corresponding scattering patterns of different modulation structures at pointing angle $\varphi=45°$, $\theta=20°$ under 10×10 metasurface array.** (a) and (b) Independently control coding sequence and corresponding 2D scattering pattern, respectively. (c) and (d)1PCM metasurfaces coding sequence and corresponding 2D scattering pattern, respectively. (e)-(f) 4PCM metasurfaces. (g)-(h) 6PCM metasurfaces. (i)-(j) 9PCM metasurfaces. (k) The sections of the scattering patterns at $\varphi=45°$ of five modulation structures.

achieve the desired beam deflection.

Take the beam pointing angle $\varphi=45°$, $\theta=20°$ as an example, the corresponding coding sequence for a 10×10 independent modulation programmable metasurface with a scale 10×10 can be calculated using equation (5), as shown in Fig. 3(a). The corresponding far-field scattering pattern calculated by equation (3) is shown in Fig. 3(b). To obtain the best beam pointing results of 10×10 BCM metasurfaces, we traverse all coding sequences to find the scattering patterns whose main-lobe beam points at $\varphi=45°$, $\theta=20°$. The optimal coding sequence and corresponding far-field scattering pattern result are shown in Figs. 3(c) and (d), respectively. Furthermore, we also obtain the coding sequences and corresponding scattering patterns for 4PCM, 6PCM, and 9PCM metasurfaces with scale 10 × 10, as shown in Figs. 3(e)-(j), respectively. Because the search space of the coding sequences is too large to traverse exhaustively, we adopt the proposed coding sequences acquisition algorithm to obtain these results. The details of the optimization process are illustrated in Supplementary Note 3. Fig. 3(k) is the sections of the scattering patterns in $\varphi=45°$ planes of five modulation metasurfaces. The theoretically calculated results demonstrate that as the number of partition items increases, the beam pointing precision increases, and the side lobe level gradually decreases, leading to a gradual improvement in angular directivity. The effectiveness and excellent performance of beam scanning in the free space of the proposed MPCM structures are also verified by the theoretically calculated results, as depicted in Supplementary Note 4.

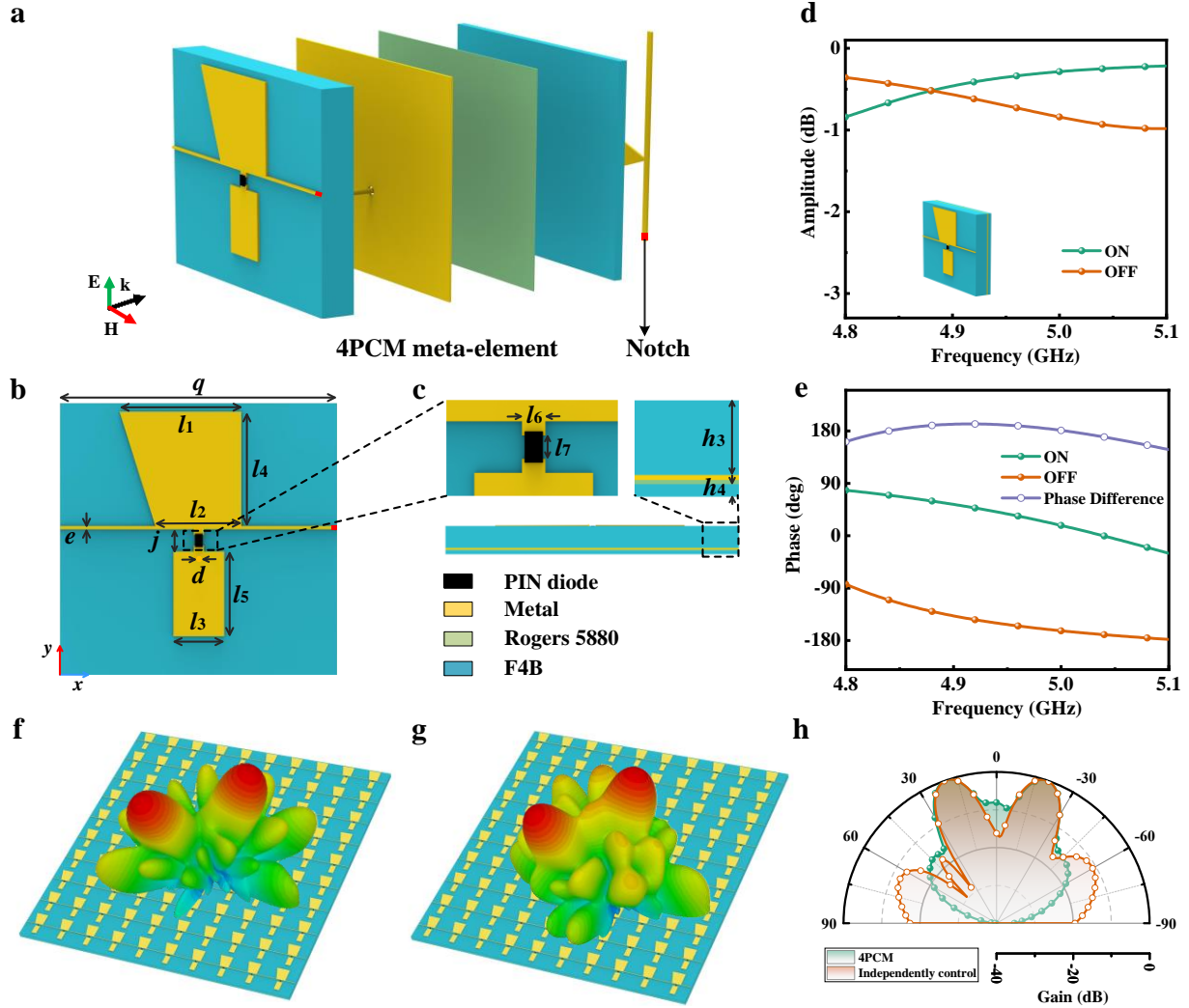

**Fig. 4 The structure and EM responses of the designed 4PCM metasurface.** (a) The 3D view of the 4PCM meta-element. (b) Front view. (c) Detailed view. The simulated results of reflected coefficient. (d) Amplitude. (e) Phase. The simulated scattering patterns of two modulation structures at pointing angle $\varphi=45°$, $\theta=20°$ under 10×10 metasurface array. (f) Independent modulation structures simulated 3D scattering pattern. (g) 4PCM structures simulated 3D scattering pattern. (h) The sections of the simulated scattering patterns at $\varphi=45°$ of two modulation structures.

## Design of the 4PCM metasurface

To validate the performance of MPCM programmable metasurfaces, we design a 1-bit phase-modulation 4PCM metasurface. The geometry diagram of the corresponding meta-element is shown in Figs. 4(a)-(c). One PIN diode is embedded on the top layer to realize the 1-bit phase-modulation ability and small notches are designed to realize the partition cross- modulation. The details of the design are depicted in the Methods section. To obtain the reflected amplitude and phase responses of the meta-element, full-wave EM simulations are performed by using the commercial EM frequency-domain solver (i.e., CST Microwave Studio 2019). The boundary conditions of the meta-element are set as "unit cell" along the $x$ and $y$ directions to mimic a two-dimensional infinite meta-element array. A $y$-polarized plane wave is normally incident upon the meta-element as the excitation. Other setups of simulation details are depicted in the Methods section. During the simulation, the on and off states of the PIN diode are equated to two lumped circuit models with specific values for the lumped components (more details in Supplementary Note 5). The simulated results are illustrated in Figs. 4(d) and (e). In the two different PIN diodes working states, the reflected phase difference is about 180° from 4.8 GHz to 5.1 GHz with about 1 dB reflection losses.

Furthermore, we simulate an independent control metasurface and a 4PCM metasurface consisting of 10×10 meta-elements with calculated coding sequences pointing at $\varphi=45°$, $\theta=20°$. The time-domain solver of CST is employed to simulate the two meta-arrays. The simulation set details are depicted in Methods. The simulated results of the two kinds of modulation metasurfaces are shown in Figs. 4(f) and (g). Fig. 4(h) shows the sections of the simulated scattering patterns in $\varphi=45°$ planes of them, which have a similar pointing angle and side lobe level. Simulated results exhibit a good agreement with the theoretically calculated results, indicating the effectiveness of the 4PCM metasurfaces in beam pointing.

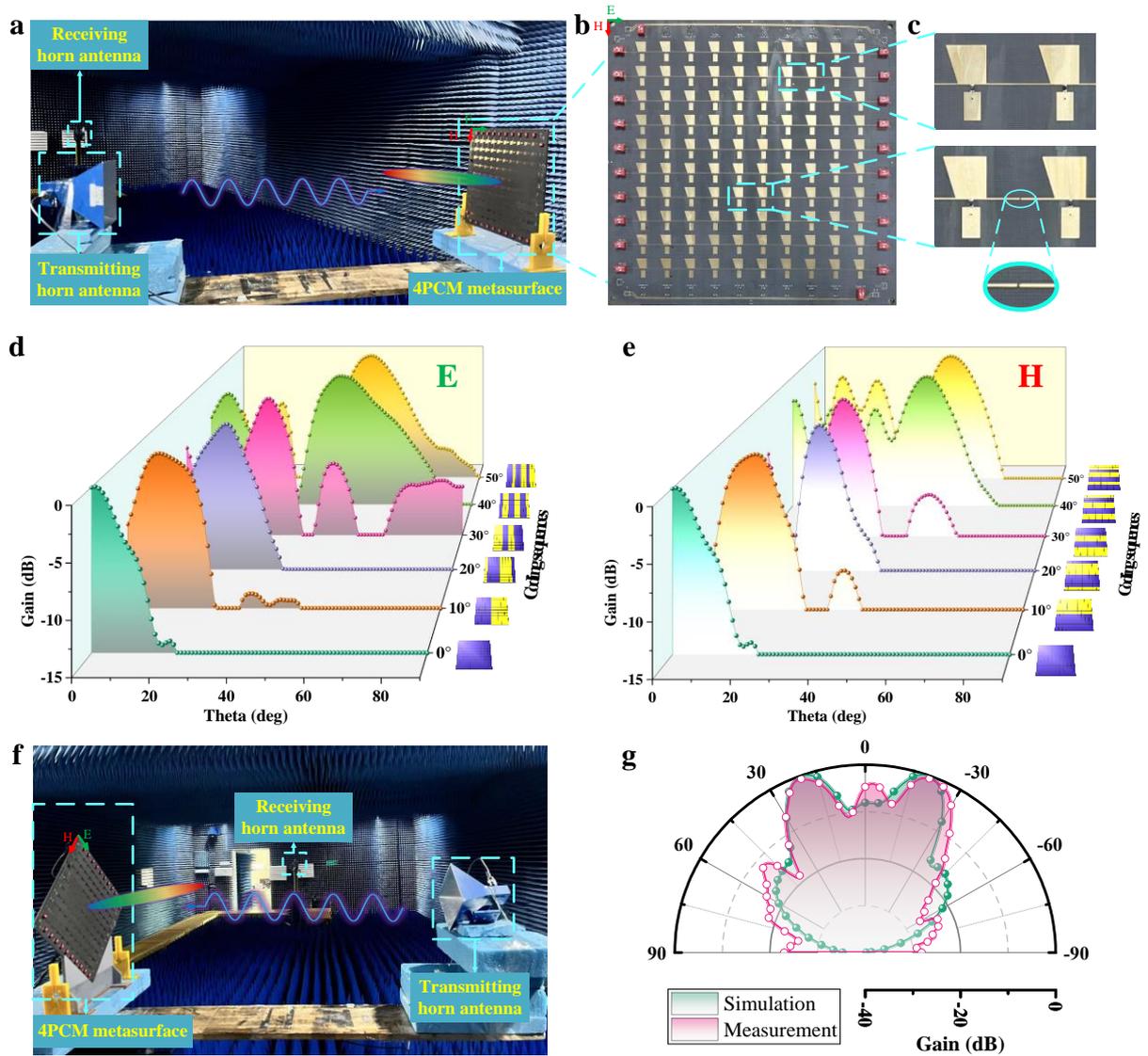

**Fig. 5 The beam scanning and pointing performance of the fabricated 4PCM metasurface at 4.95 GHz.** (a) The practical experimental environment setup for the beam scanning of the 4PCM metasurface. (b) Photograph of the manufactured 4PCM metasurface. (c) The detailed photographs of 4PCM meta-element and notches. The experimental results of beam scanning. (d) E-plane. (e) H-plane. (f) The practical experimental environment setup for the beam pointing validation. (g) The sections of the simulated and measured scattering patterns at $\varphi=45°$ plane.

## Experimental results

*Beam scanning and pointing demonstrations*

We build an experimental platform to verify the performance of the beam deflection of MPCM metasurfaces in practical applications, as shown in Fig. 5(a). We manufacture a 4PCM metasurface consisting of 10×10 periodic meta-elements using the printed-circuit-board technology. The size of the metasurface is 300×300 mm$^2$, as shown in Fig. 5(b). Some notches shown in Fig. 5(c) are placed on the metal control lines to block the DC circuits, resulting in four independent BCM structures. The dip switches placed around the array are used to quickly control the voltages of the rows and columns to achieve the target coding sequences as a demon of controllers, which can be replaced by micro control units, such as field- programmable gate arrays. The experiments are executed in the anechoic chamber with one transmitting horn antenna, one receiving horn antenna, one vector network analysis (VNA), and DC voltage sources. The details of the schematic diagram and experimental setups for beam scanning and pointing are shown in Methods and Supplementary Note 6. Specially, the practical experimental environment setup for the beam scanning in the E or H plane is shown in Fig. 5(a). We measured the beam scanning from 0° to 50° with 10° intervals in E and H planes with corresponding calculated coding sequences. The reflected beams of the 1-bit metasurfaces are normal symmetric double beams. The experimental results for elevation angles from 0° to 50° in E and H planes are shown in Figs. 5(d) and (e). The pointing angles are in good agreements with the target scanning angles in the two planes. The deviations may arise from the scale of the metasurface and the coupling of the meta-elements. Nevertheless, the results verify the beam scanning effectiveness of 4PCM metasurfaces in the E and H planes, which cannot be realized with traditional row or column modulation programmable metasurfaces.

The beam-pointing experiments are executed in the anechoic chamber with a similar set of beam-scanning experiments. The difference is that the metasurface is placed on a custom-built shelf to obtain the scattering pattern of the $\varphi=45°$ section, as shown in Fig. 5(f) (more details are given in Supplementary Note 6). We set the corresponding coding sequence with $\varphi=45°$, $\theta=20°$ as the pointing angle and run the instruments. The experimental result is contrasted with the simulated result, as shown in Fig. 5(g), which shows a good agreement with the target pointing angle. Note that the deviations may arise from the scale of the metasurface and the direct reflection from the dip switches. Nevertheless, the results in general verify the beam pointing effectiveness of 4PCM metasurfaces in the upper free space.

*Wireless communication demonstration*

Furthermore, we demonstrate the potential of the proposed 4PCM metasurface in enhancing wireless communication coverage. We take the fabricated 4PCM metasurface as the core to build an information transmission system and validate its performance of wireless communications in a realistic in-door scenario. The wireless communication experiments are conducted with a software-defined radio (SDR) platform (NI USRP-2974). The measurement schematic of the demonstration is shown in Fig. 6(a), where the 4PCM metasurface is securely mounted on a fixed disk. We use the proposed system to replace the traditional high-cost and complex radio frequency (RF) front end to connect the SDR platform. Two standard gain horn antennas operated at 4.9-6 GHz are connected to the SDR platform as the transmitter and receiver, respectively. The transmitter is positioned in front of the metasurface, aligned at the same central level. The receiver is placed at $\varphi=45°$, $\theta=20°$ in free space. We construct three scenarios to demonstrate the performance of the proposed 4PCM metasurface in wireless communication assistance: one with an inactive metasurface, i.e., i.e., the metal piece, one with a column-modulation functional metasurface (Case 2), i.e., the traditional metasurface, and one with the proposed 4PCM metasurface (Case 3), as shown in Fig. 6(b). The three cases can be realized by the 4PCM metasurface with specific voltages of row-column crossings as aforementioned. A video is converted into bitstreams and transmitted by the proposed system through the quadrature phase shift keying (QPSK) modulation method at 4.95 GHz. The SDR platform then demodulates the received signal to recover the transmitted video (for more details, please see Supporting Video 1). Figs. 6(c)-(e) show the constellation diagrams of the system when the 4PCM metasurface works in an inactive state, column-modulation state, and four partition cross-modulation state, respectively. The situations of the transmitted and received transmission video screenshots are shown in Figs. 6(f)-(g), showing that the video cannot be transmitted and experiences frame drops in Case 1 and Case 2. On the contrary, the video can only be transmitted smoothly in Case 3, which indicates the performance of the proposed 4PCM metasurface in enhancing wireless communications in free space.

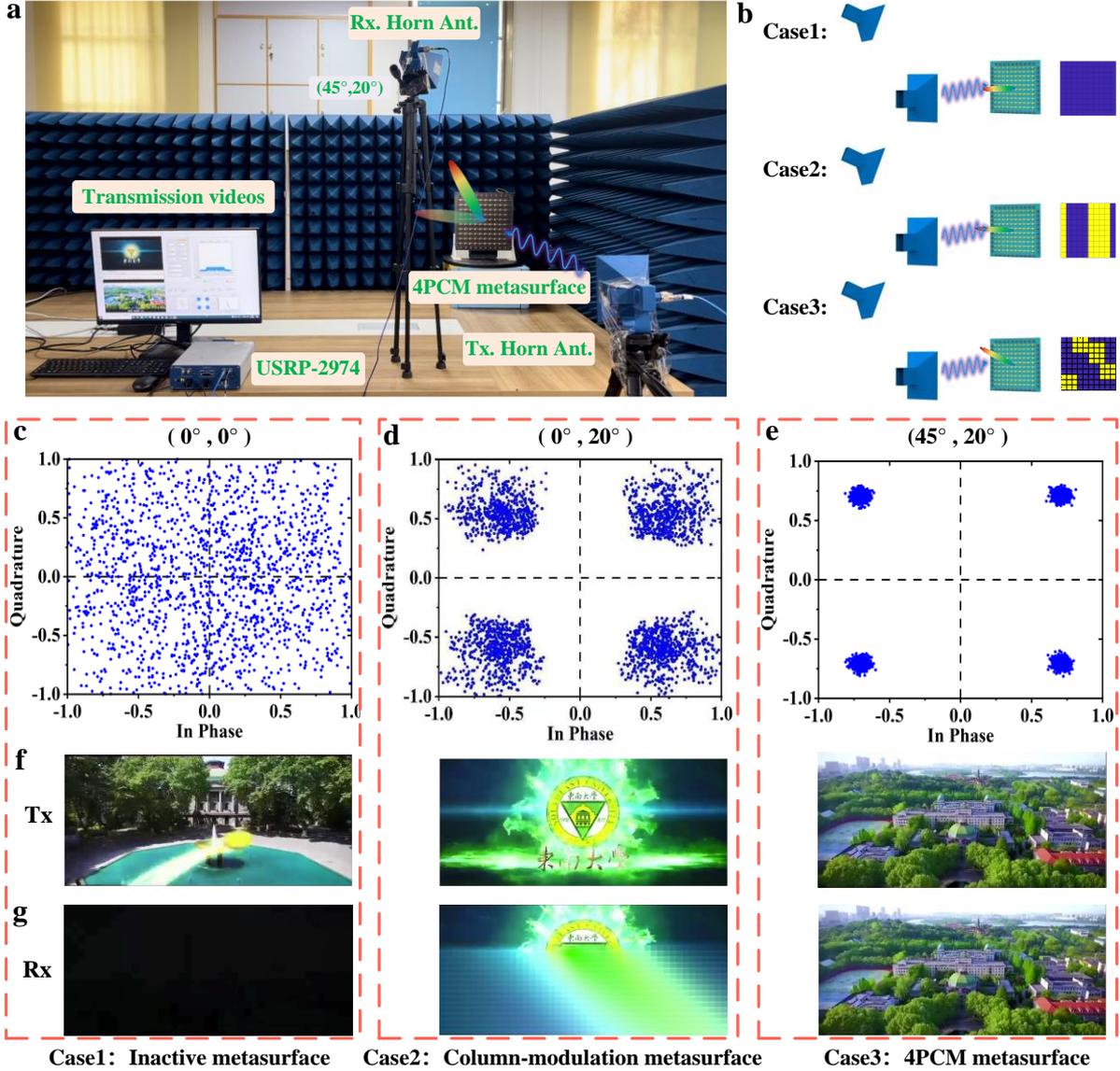

**Fig. 6 The wireless communication assistance performance of the fabricated 4PCM metasurface.** (a) The schematic diagram of the experimental setup of 4PCM metasurface in wireless communication assistance. (b) Three cases: one with an inactive metasurface, one with a column-modulation programmable metasurface, and one with the proposed 4PCM metasurface. (c)-(e) The constellation diagrams of the system in three cases. (f) The transmitted video screenshots in three cases. (g) The received video screenshots.

## DISCUSSION

We propose an MPCM programmable metasurface to balance the complexity of modulation structures with the precision of beam pointing, thereby advancing wireless communication implementation. We first formulate an innovative encoding scheme with the control voltage vectors at the row-column crossings of MPCM programmable metasurfaces. We then illustrate numerically their performance of beam pointing coverage and side-lobe suppression across various numbers of partition items in free space. A method for obtaining coding sequences is developed to achieve the target beam pointing angles with the guidance of the genetic algorithm and encoding rules. The MPCM programmable metasurfaces can realize low complexity structures with different numbers of partition items, catering to diverse precision demands across various scenarios. We design and fabricate a 4PCM programmable metasurface to verify the performance of MPCM metasurfaces. The theoretically calculated, full-wave simulated, and experimental results show good agreements, which verify the beam scanning and pointing ability in free space of the proposed MPCM metasurfaces. Additionally, we construct three cases to demonstrate its performance in a realistic indoor setting, where the 4PCM metasurface successfully transmits videos to specific receiver positions, which is an achievement beyond the capabilities of traditional programmable metasurfaces. The enhanced performance and demonstrated partition cross-modulation smartness make the MPCM metasurfaces a viable and productive new type of base station option in wireless communications.

## METHODS

### Details of the designed 4PCM metasurface

The 4PCM programmable meta-element includes five parts: the PIN diode, the quasi-H shaped metal patch, the cross-modulation structure, the dielectric, and the metal ground. The PIN diode (SMP-1320-040LF) is embedded on the top of the meta-element, manufactured by the Skyworks company. The equivalent circuit (on state with $R = 1\ \Omega$, $L = 1$ nH, $C = 1.7$ pF, and off state with $R = 8\ \Omega$, $L = 0.76$ nH, $C = 0.192$ pF) is shown in Supplementary Note 5. The metal patch is configured in a quasi-H shape with a thickness of 0.035 mm. One metal line on the patch serves to connect the positive terminal of the PIN diodes in the same row. Another metal line on the bottom of the meta-element connects the negative terminal of the PIN diodes in the same column. The dielectric substrate for the meta-element is the F4B material with $\varepsilon_r = 2.65$ and $\tan\delta = 0.001$. The adhesive layer used to bond the layers is Rogers 4450. The metal ground is a piece of copper that reflects incident EM waves. Note that two small notches are placed at the end of the feed lines to construct the MPCM metasurfaces. The detailed geometrical parameters of the 1-bit 4PCM meta-element structure are listed as follows: $q$=25 mm, $l_1$=10.9 mm, $l_2$=7.7 mm, $l_3$=4.6 mm, $l_4$=10.3 mm, $l_5$=7.7 mm, $l_6$=0.9 mm, $l_7$=0.9 mm, $e$=0.3 mm, $j$=2 mm, $d$=0.6 mm, $h_3$=3 mm, and $h_4$=0.75 mm.

### Simulations and experiments

The optimization is implemented on the platform of MATLAB 2022a. All full-wave EM simulations are executed in the commercial simulation software CST studio suite (CST Microwave Studio 2019). Both CST and MATLAB softwares are installed on a computer with an Intel (R) Core (TM) i7-9700CPU and 32 GB RAM.

The simulations of meta-elements are performed in the frequency-domain solver of CST. The boundary condition is set as "unit cell", and the electric field direction of EM waves excited by the ports is parallel to the PIN diodes. The distance between ports and meta-elements is handled using the de-embedding method. For the proposed metasurfaces, their simulations are performed in the time-domain solver of CST. The boundary condition is set as "open (add space)". The excited plane waves with the direction of electric fields parallel to the diodes illuminate the metasurfaces.

The beam scanning and beamforming experiments are conducted in an anechoic chamber. Horn antennas are employed as the transmitter and receiver. A vector network analysis (VNA, Agilent N5245A) is employed to capture the EM responses of metasurfaces controlled by a DC voltage source. The centers of horn antennas and metasurfaces are situated in the same plane while paralleling to each other. The electric field orientation of the EM waves excited by the horn antennas aligns with the diodes. To mimic planar EM wave illumination, the distance between the horn antennas and metasurfaces is larger than $2D^2/\lambda$, where $D$ is the largest size of metasurfaces and $\lambda$ is the operating wavelength. The time gating technology is adopted to mitigate the influence of scattering waves during the experiments.

## ACKNOWLEDGEMENTS

This work is supported by the National Key Research and Development Program of China (2023YFB3811502, 2018YFA0701904, 2021YFA1401002), the National Natural Science Foundation of China (62288101, 62171124, 61631007, 61571117, 61138001, 61371035, 61722106, 61731010, 11227904, 62201139, U22A2001), the National Science Fund for Distinguished Young Scholars (62225108), the 111 Project (111-2-05), the Jiangsu Province Frontier Leading Technology Basic Research Project (BK20212002), the Fundamental Research Funds for Central Universities (2242023K5002, 2242022k30004, 2242022R10055, 2242022R10185, 2242022k6003), the Jiangsu Provincial Innovation and Entrepreneurship Doctor Program, the Natural Science Foundation of Jiangsu Province (BK20220808), the Program of Song Shan Laboratory (Included in the management of Major Science and Technology Program of Henan Province, 221100211300-02, 221100211300-03), the Southeast University - China Mobile Research Institute Joint Innovation Center (R207010101125D9), and the SEU Innovation Capability Enhancement Plan for Doctoral Students (CXJH_SEU 24069).

## AUTHOR DETAILS

[1]Institute of Electromagnetic Space, Southeast University, 210096 Nanjing, China. [2]State Key Laboratory of Millimeter Waves, Southeast University, 210096 Nanjing, China. [3] School of Electronics and Communication Engineering, Guangzhou University, Guangzhou 518055, China

## AUTHOR CONTRIBUTIONS

Q.C. and T.J.C. put forward the idea and provided the instructions for this work. J.W.Z., Z.J.Q., H. D. Li., and Z.Z. conceived the designs and conducted the theoretical analyses and simulations. J.N.Z., J.Y.D, Z.Z., J.M.L., and J.W.Z. conducted parts of the designs and simulations. J.W.Z., J.Y.D., Z.X.W., S.R.W., and J.W.W. devised the measurement system, performed the experiments, and interpreted the results, with L.J.W., W.W.C., J.Y.C, and Z.H.F. assisting. J.W.Z., J.Y.D, Z.Z, J.N.Z., Q.C., X.G., and T.J.C. co-wrote the manuscript. All authors contributed to the writing and revising of the manuscript together.

## DATA AVAILABILITY

The data that support the plots within this paper and other findings of this study are available from the corresponding author upon reasonable request.

## CONFLICT OF INTEREST

The authors declare no competing interests.

# Supplementary Materials for

# Multiple-partition cross-modulation programmable metasurface empowering wireless communications

Jun Wei Zhang *et al*.

**This PDF file includes:**

1. The encoding rules of control voltages and PIN diode working states
2. Guidance of partitioning and the size of sub-BCM metasurfaces determination
3. Details of the optimization process
4. Beam scanning performance of MPCM metasurfaces in theoretically calculations
5. Extraction of the equivalent circuit values of PIN diodes
6. Schematic diagram and the experimental setups for beam scanning and pointing of programmable metasurfaces

Figs. S1 to S8

**Note 1: Encoding rules of control voltages and PIN diode working states**

The PIN diodes' working states of the basic cross-modulation (BCM) metasurfaces are determined by the control voltages of the row-column crossings, as shown in Figs. S1(a) and (b). We define vectors $V$ with $M$ elements and $H$ with $N$ elements to represent the control voltages of all rows and columns, respectively. Further let $V^i = 1$ and $V^i = 0$ ($i = 0, 1, …, M$) represent high and low voltage at the $i$ row, respectively. Similarly, let $H^j = 0$ and $H^j = 1$ ($j = 0, 1, …, N$) represent high and low voltage at the $j$ column, respectively. For example, when $V = [0, 0, 0, …, 0]$, all rows are at low voltages. When $H = [0, 0, 0, …, 0]$, all rows are at high

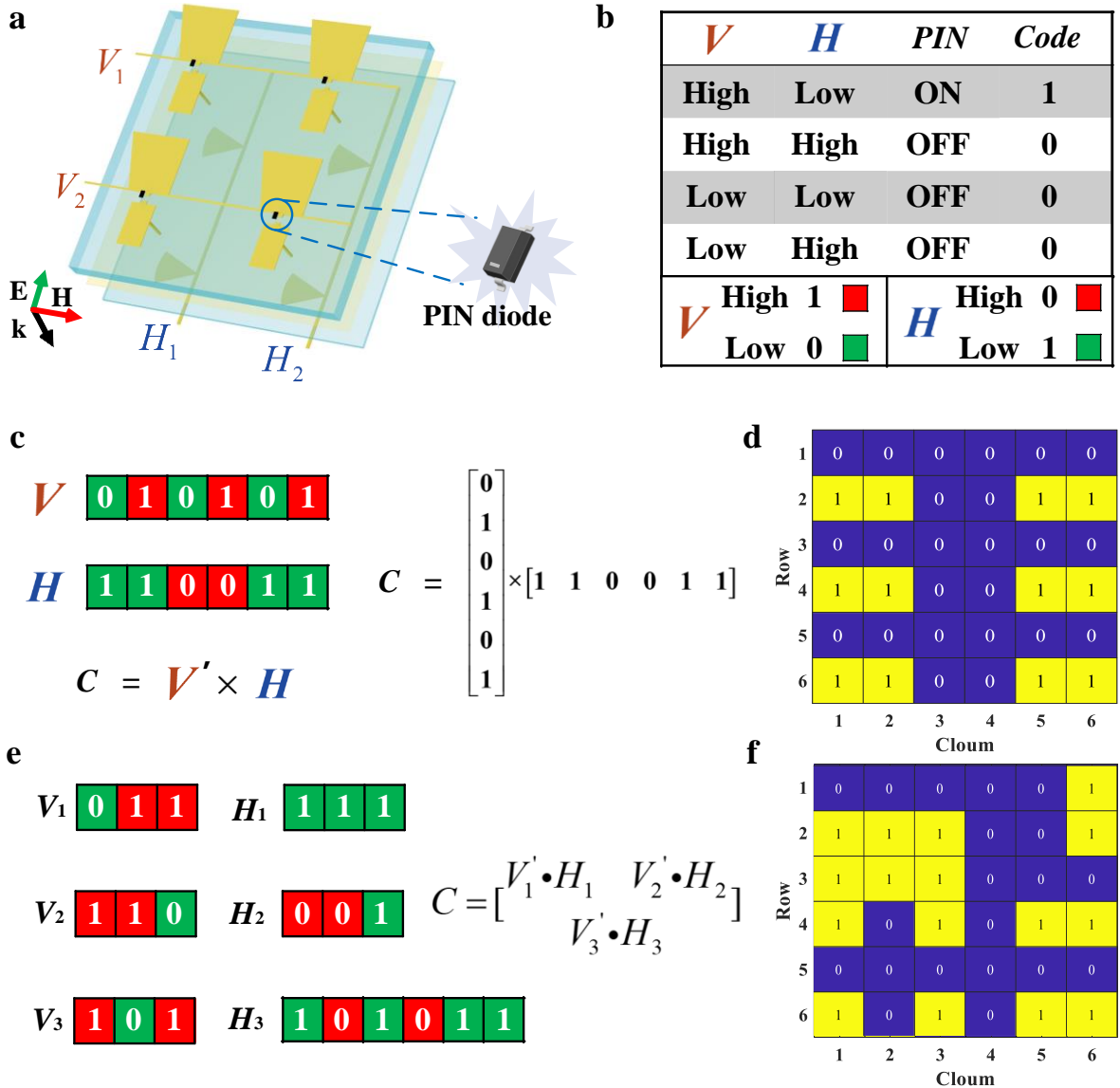

**Fig. S1 Encoding rules of MPCM metasurfaces.** (a) The placement of the row-column crossings and PIN diodes. (b) The coding rules of BCM method and the control codes of row-column crossings. (c) The encoding example of a BCM metasurface with a scale of 6×6 and the coding sequences calculated formula. (d) The corresponding coding sequences of the BCM metasurface. (e) The encoding example of a 3PCM metasurface with a scale of 6×6 and the calculated formula. (f) The corresponding coding sequences of the 3PCM metasurface.

voltages. According to the two vectors defined above, we can obtain matrix $C$

$$C = V^T \cdot H, \qquad (1)$$

where $C$ represents the working states of the diodes in the metasurface array. The element $C^{(m,n)}$ ($m = 1,2,\ldots, M$ and $n = 1,2,\ldots, N$) represents the working state of the PIN diode located at the $m$-th row and $n$-th column, where $C^{(m,n)} = 0$ and $C^{(m,n)} = 1$ represents off and on states of the ($m$, $n$)-th diode, respectively.

For example, a BCM metasurface with the scale of 6×6 contains six row control terminals and six column control terminals, as shown in Fig. S1(c). Assuming that $V = [0, 1, 1, 1, 0, 1]$ and $H = [1, 0, 1, 0, 0, 1]$, the working states of PIN diodes in the metasurface are

$$C = V^T \cdot H = \begin{bmatrix} 0 \\ 1 \\ 1 \\ 1 \\ 0 \\ 1 \end{bmatrix} \cdot [1 \; 0 \; 1 \; 0 \; 0 \; 1] = \begin{bmatrix} 0 & 0 & 0 & 0 & 0 & 0 \\ 1 & 0 & 1 & 0 & 0 & 1 \\ 1 & 0 & 1 & 0 & 0 & 1 \\ 1 & 0 & 1 & 0 & 0 & 1 \\ 0 & 0 & 0 & 0 & 0 & 0 \\ 1 & 0 & 1 & 0 & 0 & 1 \end{bmatrix}. \qquad (2)$$

The coding sequence is shown in Fig. S1(d). Generally, the digital coding metasurfaces employ the codes "0" and "1" to represent the opposite EM phase responses (0 and π). Here, the codes "0" and "1" correspond to the off and on states of diodes, respectively. Therefore, the phase profile of BCM metasurfaces can be represented by the matrix πC.

For a multiple partition cross-modulation (MPCM) programmable metasurface, we assume that it consists of $K$ BCM structures. Accordingly, we define $V_1, V_2, \ldots, V_K$, and $H_1, H_2, \ldots, H_K$ to represent the vertical and horizontal voltage codes of the $K$ parts, respectively. The working states of PIN diodes in each BCM structure, i.e., $C_1, C_2, \ldots, C_K$ can be calculated by equation (1). Through the two group vectors and matrix operations, the PIN diode working states of $K$ partition cross-modulation metasurfaces ($C_{KPCM}$) can be represented by

$$C_{KPCM} = \begin{bmatrix} C_1 & C_2 & \cdots \\ \vdots & \vdots & \vdots \\ \cdots & C_{K-1} & C_K \end{bmatrix} = \begin{bmatrix} V_1^T \cdot H_1 & V_2^T \cdot H_2 & \cdots \\ \vdots & \vdots & \vdots \\ \cdots & V_{K-1}^T \cdot H_{K-1} & V_K^T \cdot H_K \end{bmatrix}. \qquad (3)$$

Consider a three partition cross-modulation (3PCM) metasurface with a scale of 6×6, which consists of two 3×3 BCM metasurfaces and one 3×6 BCM metasurface. The control voltages of their rows and columns are $V_1 = [0, 1, 1]$, $V_2 = [1, 1, 0]$, $V_3 = [1, 0, 1]$, $H_1 = [1, 1, 1]$, $H_2 = [0, 0, 1]$, and $H_3 = [1, 0, 1, 0, 1, 1]$, respectively, as shown in Fig. S1(e). The working states of the PIN diodes in this metasurface are

$$C_{3PCM} = \begin{bmatrix} V_1^T \cdot H & V_2^T \cdot H_2 \\ V_3^T \cdot H_3 \end{bmatrix} = \begin{bmatrix} \begin{bmatrix} 0 \\ 1 \\ 1 \end{bmatrix} \cdot [1 \ 1 \ 1] & \begin{bmatrix} 1 \\ 1 \\ 0 \end{bmatrix} \cdot [0 \ 0 \ 1] \\ \begin{bmatrix} 1 \\ 0 \\ 1 \end{bmatrix} \cdot [1 \ 0 \ 1 \ 0 \ 1 \ 1] \end{bmatrix} = \begin{bmatrix} 0 & 0 & 0 & 0 & 0 & 1 \\ 1 & 1 & 1 & 0 & 0 & 1 \\ 1 & 1 & 1 & 0 & 0 & 0 \\ 1 & 0 & 1 & 0 & 1 & 1 \\ 0 & 0 & 0 & 0 & 0 & 0 \\ 1 & 0 & 1 & 0 & 1 & 1 \end{bmatrix}. \quad (4)$$

The coding sequence is shown in Fig. S1(f). Correspondingly, the phase profile of the programmable metasurface can be represented by $\pi C_{3PCM}$.

**Note 2: Guidance of partitioning and the size of sub-BCM metasurfaces determination**

For an MPCM programmable metasurface with a scale of $M\times N$, there are $M+N$ row and column control terminals when its partition item number is one, i.e., BCM metasurfaces. When the number of partition items are $K$, i.e., $K$ partition cross-modulation (KPCM), there are many partitioning cases, resulting in various coding sequences. For the sake of simplicity, we use the equalization strategy to separate the sub-BCM metasurface, ensuring that each piece is essentially the same size. For example, assuming four partition cross-modulation (4PCM) programmable metasurface with the scale of 10×10. As shown in Figs. S2(a)-(c), the size of the sub-BCM metasurfaces can be 3×3, 7×7, 4×5, 6×3, 5×5, and 6×7 etc. Here, we adopt the strategy of equalization to verify the effectiveness of the proposed MPCM method. The 4PCM programmable metasurface is separated into four BCM metasurfaces with a size of 5×5, as shown in Fig. S2(c). This partitioning method may not obtain the best coding sequences for the desired far-field scattering patterns. Nevertheless, we still can use the classic equalization strategy to verify the performance of MPCM metasurfaces. The advanced separation strategies can be further investigated to obtain higher performance.

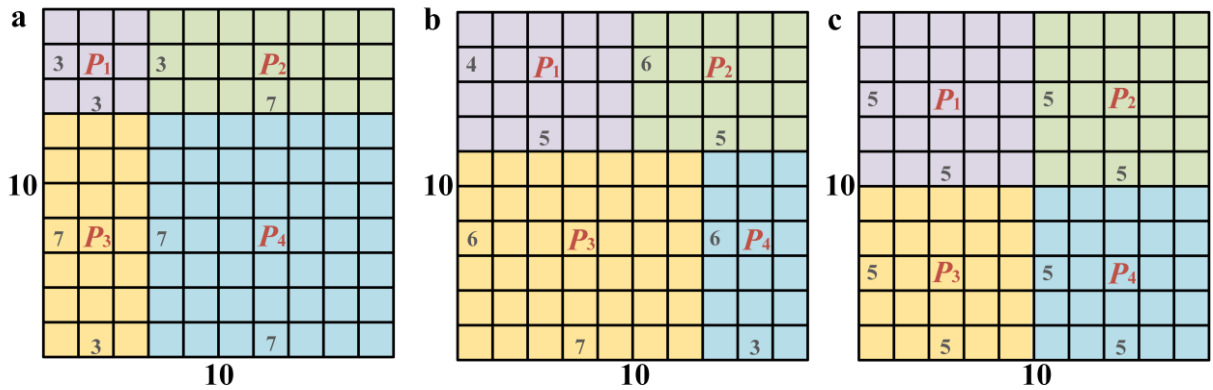

**Fig. S2 Different partitioning methods of a 4PCM metasurface with the scale of 10×10.** (a) The sizes of the sub-BCM metasurfaces are 3×3, 7×7, and 7×3. (b) The sizes of the sub-BCM metasurfaces are 4×5, 6×5, 6×7, and 6×3. (c) The sizes of the sub-BCM metasurfaces are all 5×5, which are separated with the equalization strategy used in this work.

**Note 3: Details of the optimization process**

Since the novel row and column crossings are coupled with each other, the traditional methods of obtaining coding sequences are not suitable. The coding sequence acquisition is a non-convex problem. Therefore, we employ the classic heuristic algorithm, i.e., genetic algorithm (GA), to select the best coding sequences for desired far-field scattering patterns. We set the control voltages of the row-column crossings as the input parameters, and the phase profile of the metasurfaces' beam pointing at desired angles as the optimization targets.

Specifically, we first calculate the phase profile of the metasurfaces' beam pointing at the main-lobe direction angle ($\varphi_0$, $\theta_0$) according to the following formula

$$\Phi(m,n) = -k \cdot (x_i \cdot \sin\theta_0 \cdot \cos\varphi_0 + y_i \cdot \sin\theta_0 \cdot \sin\varphi_0), \tag{5}$$

where $\Phi(m, n)$ is the phase of the $(m, n)$-th meta-element, and $(x_i, y_i)$ represents its position in the Cartesian coordinate system. $\theta_0$ and $\varphi_0$ represent the target main-lobe angle of elevation and azimuth, respectively.

We then define a vector $X = [V_1, V_2, \ldots, V_K, H_1, H_2, \ldots, H_K]$, which includes all the control voltage terminals of MPCM metasurfaces, i.e., the vectors $V_1, V_2, \ldots, V_K$ and $H_1, H_2, \ldots, H_K$. The distribution reflection phases of MPCM metasurfaces can be described as $\pi C_{KPCM}$, calculated by equation (3), which is a function of $X$. Assuming that the beam deflection angle is ($\varphi_0$, $\theta_0$). We define the corresponding phase distribution of the programmable metasurfaces as $\Phi$, calculated by equation (5). Then we can define the objective function for the coding sequences optimization as follows

$$U_1(X) = \frac{\|\pi C_{KPCM}(X) - \Phi\|_F}{M \cdot N}, \tag{6}$$

where $U_1$ represents the objective function evaluated at given values of $X$, $\|\cdot\|_F$ represents the Frobenius norm of a given matrix. The symbol $\Phi$ is used here to assist the metasurface in achieving the desired main-lobe direction angles. Based on equation (6), the optimization problem can be formulated by

$$X^* = \arg\min_{X} U_1(X), \tag{7}$$

where $X^*$ represents the optimized solution of $X$, containing the voltages of rows and columns that can achieve the desired beam deflection.

For example, we assume that a 4PCM programmable metasurface with a scale of 10×10 is separated into four BCM metasurfaces with the sizes of 5×5. There are twenty control terminals of row-column crossings with high or low voltages, which can generate $2^{20}$ number of coding sequences. The vector $X$ has twenty elements. Assuming that the beam pointing angle of the 4PCM metasurfaces is $\varphi=45°$, $\theta=20°$, the corresponding phase profile $\Phi$ can be calculated by equation (5). The objective function value $U_1$ can be represented by the $X$ and $\Phi$. We then select the GA with the setup of forty iterations, each with fifty particles to obtain the optimized coding sequences. The convergence of the optimization process of obtaining the voltages of rows and columns is shown in Fig. S3(g). In the entire optimization process, the optimization framework takes twenty iterations to converge to a good solution within five minutes. Correspondingly, the convergence of the optimization processes of BCM, six partition cross-modulation (6PCM), and nine partition cross-modulation (9PCM) metasurfaces are shown in Figs. S3(f), (h), and (i), respectively. The lowest objective function values of MPCM structures decreases with the increase of the number of partition items are shown in Fig. S3(j), indicating the effectiveness of the proposed MPCM metasurfaces. The optimized coding sequences and corresponding far-field scattering patterns of the five modulation metasurfaces are shown in Figs. S3(a)-(e), which has a significant improvement of beam pointing precision with the increase of the number of partition items.

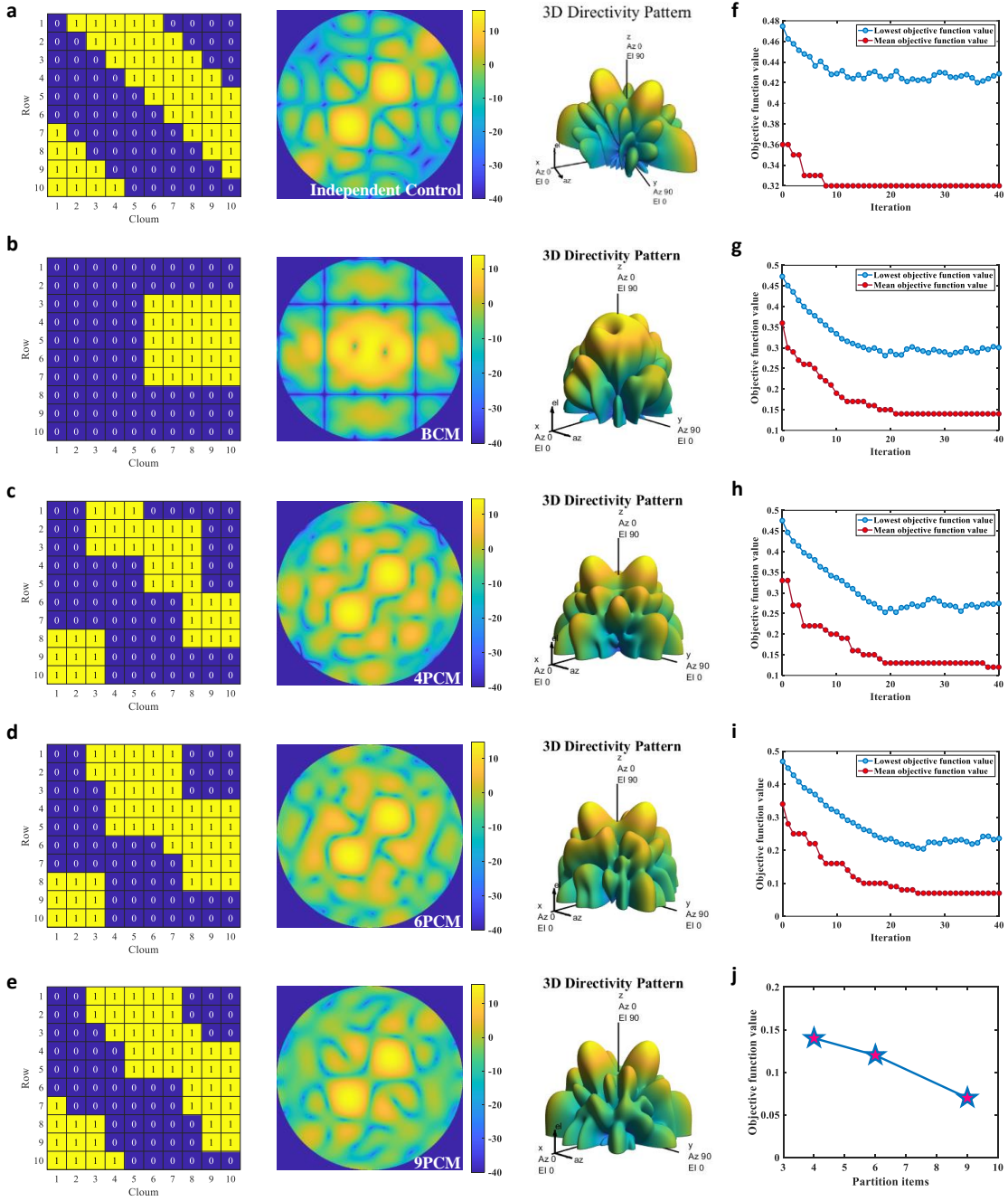

**Fig. S3 The theoretically calculated beam pointing results of programmable metasurfaces with the scale of 10×10 in different modulation methods at 4.95 GHz.** (a) The coding sequences, far-field scattering patterns at main-lope angle $\varphi=45°$, $\theta=20°$ with the independent control method. (b) The optimized coding sequences, far-field scattering patterns at main-lope angle $\varphi=45°$, $\theta=20°$ with the BCM method. (c) The 4PCM method. (d) The 6PCM. (e) The 9PCM. (f) The convergence of the optimization process with BCM method. (g) The 4PCM method. (h) The 6PCM method. (i) The 9PCM method. (j) Lowest objective function values of the four kinds of MPCM methods.

**Note 4: Beam scanning performance of MPCM metasurfaces in theoretically calculations**

The proposed MPCM programmable metasurfaces have the beam scanning ability in free space.

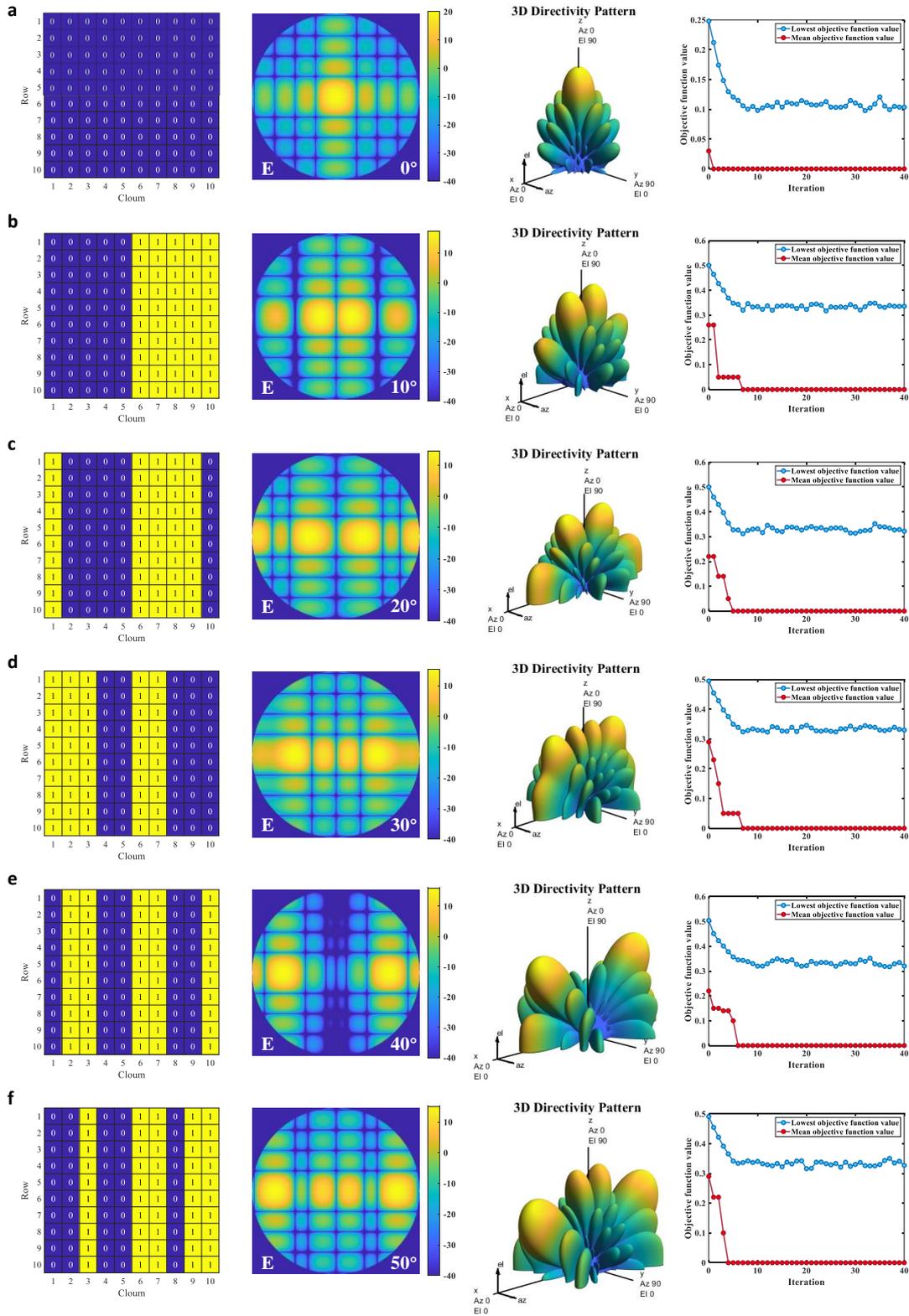

**Fig. S4 The theoretically calculated beam scanning results of BCM metasurfaces with the scale of 10× 10 in E plane.** (a) The coding sequences, far-field scattering patterns and corresponding optimization process of $\theta=0°$. (b) $\theta=10°$. (c) $\theta=20°$. (d) $\theta=30°$. (e) $\theta=40°$. (f) $\theta=50°$.

When the voltages of column control terminals are all low, the MPCM metasurfaces can be equivalent to row-modulation metasurfaces which can realize the beam scanning in the E plane.

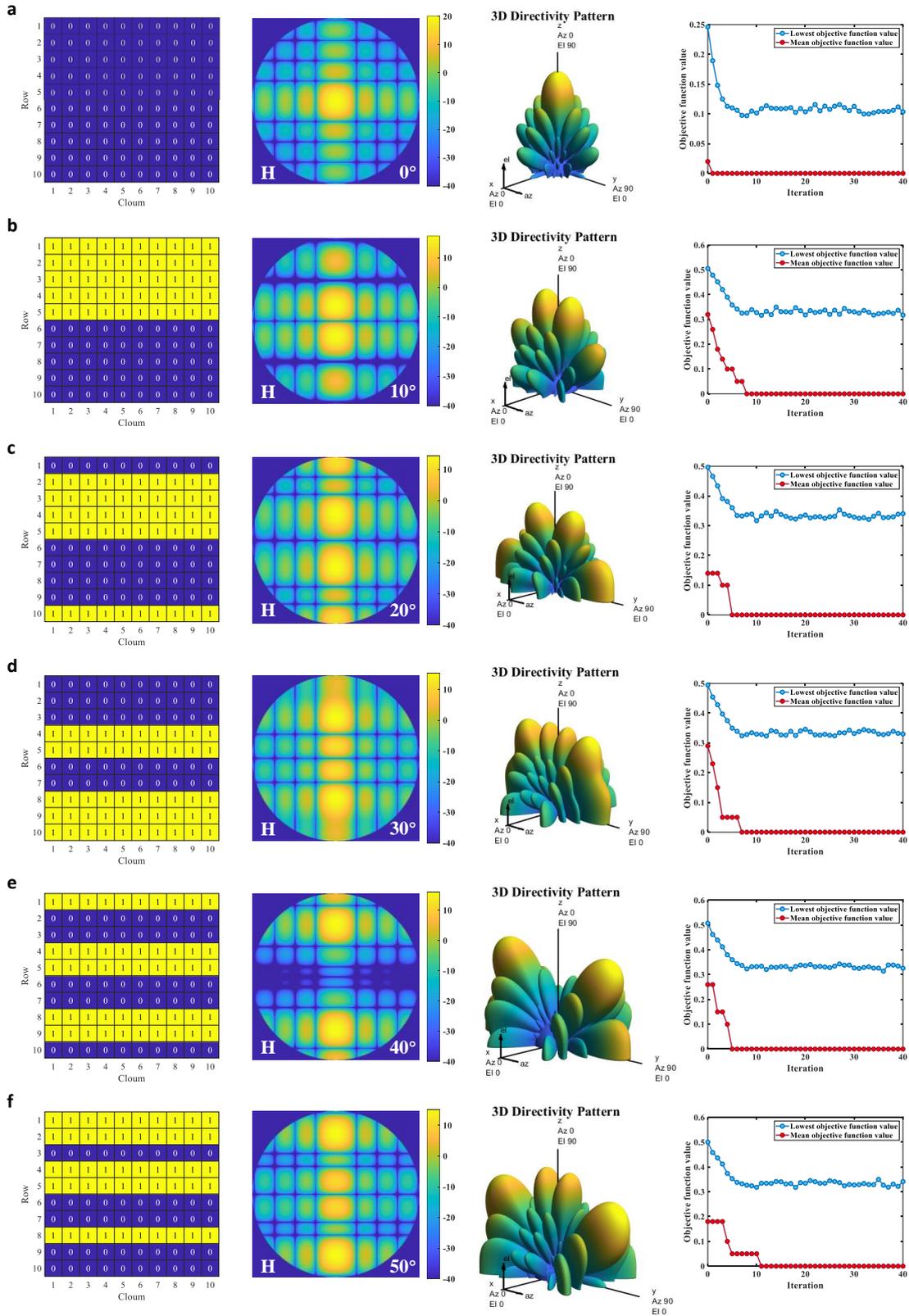

**Fig. S5 The theoretically calculated beam scanning results of BCM metasurfaces with the scale of 10× 10 in H plane.** (a) The coding sequences, far-field scattering patterns and corresponding optimization process of $\theta=0°$. (b) $\theta=10°$. (c) $\theta=20°$. (d) $\theta=30°$. (e) $\theta=40°$. (f) $\theta=50°$.

Additionally, when the voltages of the row control terminals are all high, the MPCM metasurfaces can be equivalent to column-modulation metasurfaces which can realize the beam scanning in the H plane. Hence, the beam scanning ability can be easily realized by BCM metasurfaces without any physical assistance, which cannot be realized by the traditional row or column-modulation metasurfaces. Furthermore, we use the proposed optimization method to obtain twelve group coding sequences of a 10×10 BCM programmable metasurface, whose main-lobe angle points at $\varphi=0°$ and $90°$, $\theta$ from $0°$ to $50°$ with $10°$ intervals, respectively. The corresponding coding sequences and convergences of the optimization processes are shown in Figs. S4 and S5. Based on the coding sequences, we can obtain the corresponding amplitude and phase profiles of metasurfaces. According to the convolution operations of coding metasurfaces, the relationship between the scattering patterns and phase distribution ($\pi C$) of these metasurfaces can be expressed by

$$F(\theta,\varphi) = G(u,v) = \mathcal{F}\{Ae^{j\pi C}\}, \tag{8}$$

where $u = \sin\theta \cdot \cos\phi$, $v = \sin\theta \cdot \sin\phi$, $C$ is the coding sequences, and $A$ is the amplitude profiles.

The far-field scattering patterns of these coding sequences can be obtained, as shown in Figs. S4 and S5, whose beams are pointed precisely at the target angles. The theoretically calculated results verify the advancement of the proposed MPCM programmable metasurfaces in beam scanning.

**Note 5: Extraction of the equivalent circuit values of PIN diodes**

The equivalent circuit models selected in this work are the basic circuits according to the working mechanism of the diodes, as shown in Fig. S6(a). $R$, $L$, and $C$ represent the values of the resistor, inductor, and capacitor in the equivalent circuit model, respectively. $Z$ represents the corresponding equivalent impedance of the circuit model, which can be calculated by the following formula

$$Z = R + j(\frac{4\pi^2 \cdot f^2 \cdot L \cdot C - 1}{2\pi \cdot f \cdot C}). \tag{9}$$

We obtain the actual $Z$ impedance of the PIN diodes at on or off states based on the diodes' S-parameters extracted from the data sheet or measurement results. We then use the numerical fitting method to fit the two impedances for obtaining equivalent model parameter values. The fitting results (at on state, $R = 1\ \Omega$, $L = 1$ nH, $C = 1.7$ pF, and at off state $R = 8\ \Omega$, $L = 0.76$ nH, $C = 0.192$ pF from 4 GHz to 6 GHz band) of the PIN diodes are shown in Figs. S6(b) and (c). The process is quick, but the limitation is that this method does not have broadband properties. For the application of broadband metasurfaces, the segmented numerical fitting method can be employed to address this issue.

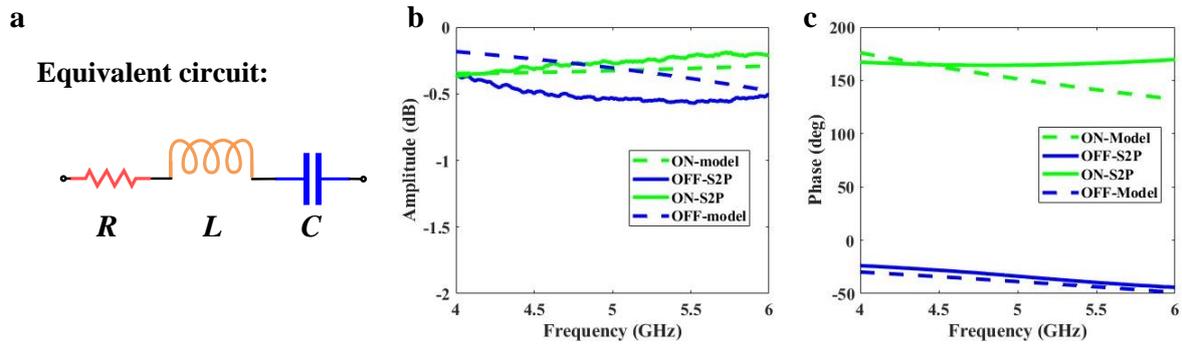

**Fig. S6 The extraction process of the equivalent circuit values of PIN diodes.** (a) The basic equivalent circuit model of the PIN diodes working at on or off state. (b) The amplitude responses of $S_{11}$ of the PIN diodes in the S2P files and the fitting equivalent circuit values at on and off working states. (c) The phase responses.

**Note 6: Schematic diagram and the experimental setups for beam scanning and pointing of MPCM metasurfaces**

The beam scanning and beamforming experiments are conducted in an anechoic chamber. The schematic diagram of the experimental setup for beam scanning is shown in Fig. S7(a). Two standard horn antennas are employed as the transmitter and receiver. A vector network analysis (VNA, Agilent N5245A) is used to capture the EM responses of the fabricated metasurfaces fed by a DC voltage source. An antenna measured system is utilized to obtain the far-field scattering patterns of the fabricated 4PCM metasurface with the aforementioned microwave devices. The centers of horn antennas and metasurfaces are situated in the same plane while paralleling to each other. The electric field orientation of the EM waves excited by the horn antennas aligns with the diodes. To mimic planar EM wave illumination, the distance between the horn antennas and metasurfaces is set to be larger than $2D^2/\lambda$, where $D$ is the largest size of metasurfaces and $\lambda$ is the operating wavelength. The practical experimental distances between horn antennas and

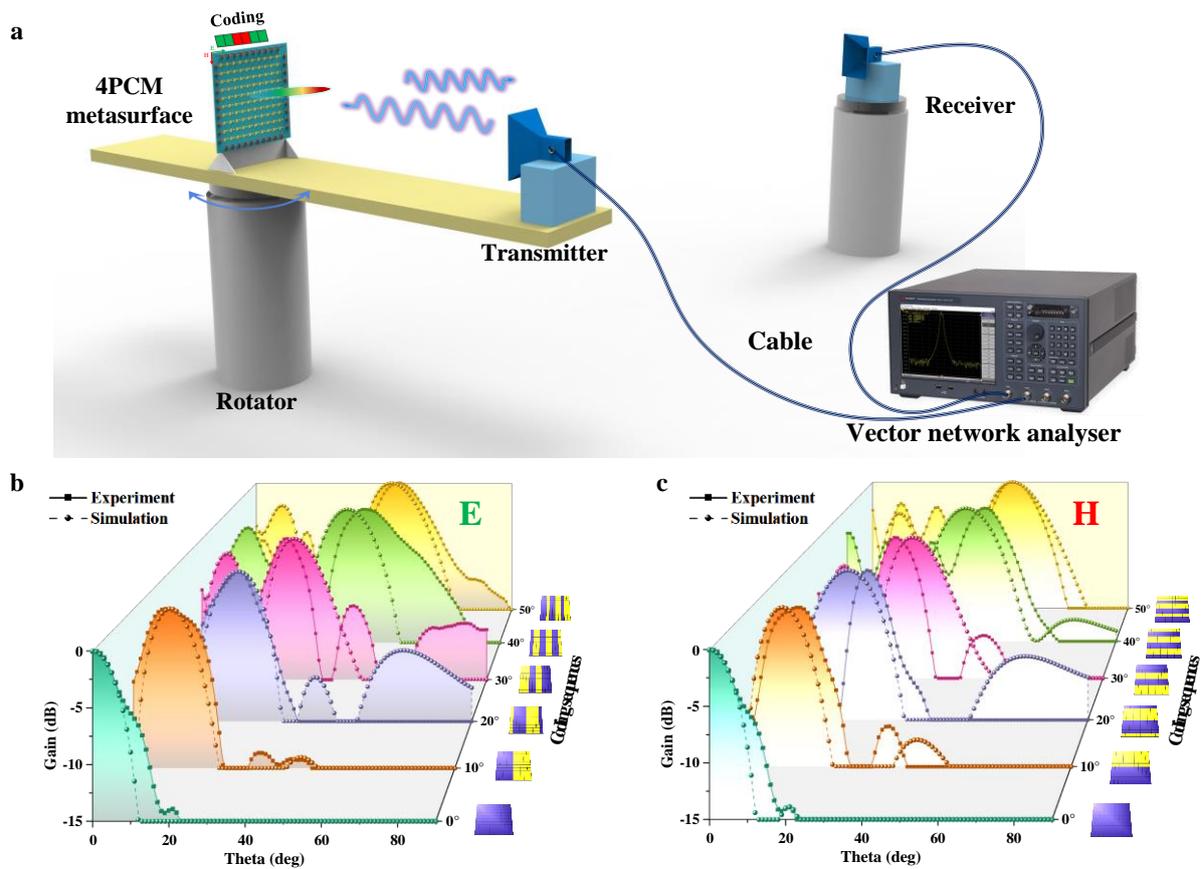

**Fig. S7** (a) The schematic diagram of the experimental setup of the fabricated 4PCM metasurface beam scanning in E and H planes. (b) The comparison between simulated and measured results of beam scanning from 0° to 50° in the E plane. (c) The H Plane.

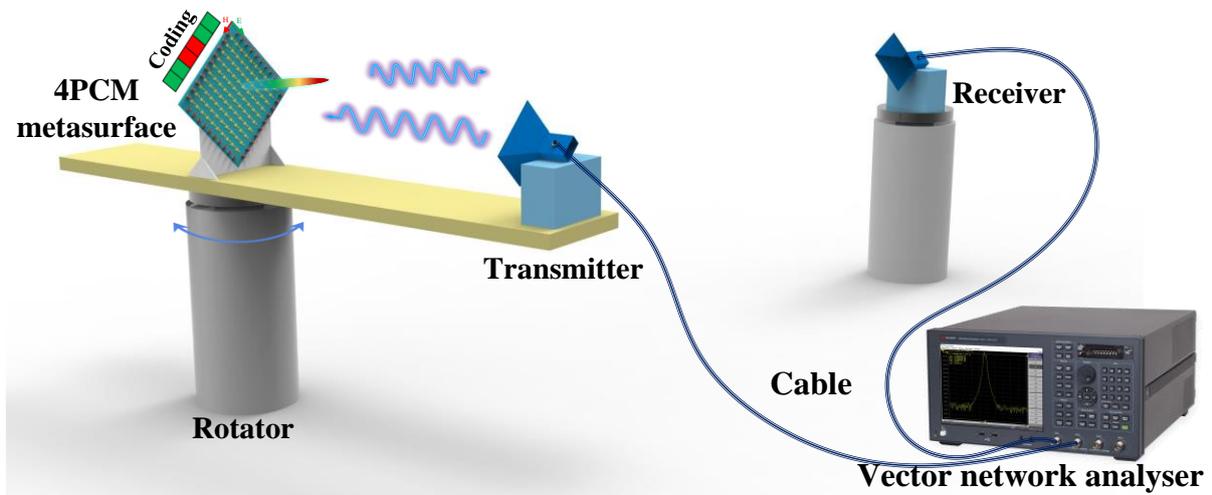

**Fig. S8** The schematic diagram of the experimental setup of the fabricated 4PCM metasurface beam pointing

metasurfaces are 2 m and 5 m, respectively. The time gating technology is adopted to mitigate the influence of scattering waves during the experiments.

During the experimental process of beam scanning in E and H planes. The 4PCM metasurface is placed vertically and horizontally on the platform to enable the scanning of $\varphi=0°$ and 90° planes, i.e., E and H planes in turn. When scanning the E plane, the voltages of columns are all low. The corresponding beam scanning coding sequences are converted to the voltages of rows by using the DC voltage source. By tuning the ON and OFF of the selected dip switches, the target voltages of the rows and columns can be realized. Note that the high voltage used here is 0.9 V, while the low is 0 V. When scanning the H plane, the voltages of rows are all high. The voltage codes of columns are the vertical coding sequences. The rotator rotates from -90° to 90°, i.e., $\theta \in [-90°, 90°]$. The simulated and measured results operating at 4.95 GHz are shown in Figs. S7(b) and (c). The deviation may arise from the fabricated tolerances and the "edge truncation effect" of the fabricated metasurface. Nevertheless, the results with the agreement can verify the performance of the beam scanning ability.

When measuring the beam pointing performance of the 4PCM metasurface in free space, due to the beam pointing angle $\varphi=45°$, $\theta=20°$, the metasurface is placed on a specially designed shelf to allow the metasurface to be placed perpendicular to the $\varphi=45°$ plane. The corresponding schematic diagram is shown in Fig. S8. The voltages of corresponding rows and columns can be obtained from the coding sequences and coding rules. The measured results are shown in Fig. 5(g), demonstrating the effectiveness of the proposed MPCM metasurfaces in real applications.